\documentclass[aps,prl,twocolumn,groupedaddress]{revtex4-1}

\usepackage{amsmath}
\usepackage{bm}
\usepackage{verbatim}
\usepackage{graphicx}
\usepackage{subfigure}
\usepackage{rotating}

\usepackage{psfrag}
\usepackage{epsfig}
\usepackage{lpic}

\newcommand{\ffrac}[2]{\mbox{$\frac{#1}{#2}$}}
\def\half{\mbox{$\frac{1}{2}$}} 
\def\quarter{\mbox{$\frac{1}{4}$}} 
\def\eighth{\mbox{$\frac{1}{8}$}}

\newcommand{\eq}[1]{\begin{align}#1\end{align}}

\def\p{\partial}
\def\zbar{\bar{z}}
\def\sigmab{\bm{\hat{\sigma}}}

\def\rhob{\bm{\rho}}
\def\rhobt{\bm{\hat{\rho}}}

\def\psibt{\bm{\hat{\psi}}}
\def\rb{\bm{r}}
\def\fb{\bm{f}}
\def\lb{\bm{\ell}}
\def\ltildeb{\tilde{\bm{\ell}}}
\def\stildeb{\tilde{\bm{s}}}
\def\epsb{\bm{\hat{\varepsilon}}}
\def\delb{\bm{\hat{\delta}}}

\newcommand{\mm}[0]{\mbox{$-$}\!\!}

\newcommand{\ds}{\mbox{d}^*\!}

\def\half{\mbox{$\frac{1}{2}$}} 
\def\quarter{\mbox{$\frac{1}{4}$}} 
\def\eighth{\mbox{$\frac{1}{8}$}}

\def\lgc{\bm{\ell}_g^c}
\def\lgcp{\bm{\ell}_g^{c^+}}
\def\lgcm{\bm{\ell}_g^{c^-}}
\def\fgc{\bm{f}_g^c}
\def\sgct{\bm{s}_{gc}^t}

\def\tr{\mbox{tr}}

\begin{document}


\title{Laws of Granular Solids. Geometry and Topology}


\author{Eric DeGiuli}
\affiliation{Department of Mathematics, University of British Columbia}
\author{Jim McElwaine}
\affiliation{Department of Applied Mathematics and Theoretical Physics, University of Cambridge}


\date{\today}

\begin{abstract}
In a granular solid, mechanical equilibrium requires a delicate balance of forces at the disordered grain scale. To understand how macroscopic rigidity can emerge in this amorphous solid, it is crucial that we understand how Newton's laws pass from the disordered grain scale to the laboratory scale. In this work, we introduce an exact discrete calculus, in which Newton's laws appear as differential relations at the scale of a \textit{single grain}. Using this calculus, we introduce gauge variables which describe identically force- and torque-balanced configurations. In a first, intrinsic formulation, we use the topology of the contact network, but not its geometry. In a second, extrinsic formulation, we introduce geometry with the Delaunay triangulation. These formulations show, with exact methods, how topology and geometry in a disordered medium are related by constraints. In particular, we derive Airy's expression for a divergence-free, symmetric stress tensor in two and three dimensions.
\end{abstract}


\maketitle

\section{\label{sect:intro}Introduction}
Poised between fluids and solids, granular media pose outstanding challenges in fundamental physics \cite{Jaegeretal:1996,*LiuNagel:1998,*Catesetal:1998}. At finite pressure, a cohesionless, frictional granular medium resists bulk and shear deformation, and is therefore a solid. However, application of a sufficiently large shear stress will cause the medium to deform indefinitely. It is largely unknown how grain-scale quantities vary with the applied stress, and therefore we cannot yet predict the yield stress from first principles. 







At the grain scale, the mechanical stability of the solid phase requires a delicate balance of forces and geometry. For nearly rigid grains, like sand, a good approximation is that intergranular forces are transmitted at discrete contacts. The positions of these contacts, and the forces across them, must be arranged so that the net force and net torque on each grain vanish. To understand how a solid granular medium responds to a macroscopically applied stress, we need to understand how these ``microscopic'' constraints pass from the disordered grain scale to the laboratory scale.

To provide insight on this process, in this note we show how forces and torques can be derived from discrete potentials which identically satisfy these constraints. These potentials have continuous analogs which can be exactly related to grain-scale quantities by a discrete calculus, which we develop. In particular, we consider a static, frictional packing $\Omega$ in dimensions $d=2$ and $d=3$, in the absence of body forces. Each of the $N$ soft, repulsive, identical disks ($d=2$) or spheres ($d=3$) $g$ is subject to force and torque balance:
\begin{equation}
\label{Newton}
0 = \sum_{c\in C^g} \fgc, \;\;\; 0 = \sum_{c\in C^g} (\rb^c - \rb^g) \times \fgc.
\end{equation}
Here $\fgc$ is the contact force exerted on grain $g$ at contact $c$, $\rb^c$ is the position of $c$, and $\rb^g$ is position of the center of $g$. Each contact force must also satisfy a repulsive condition $(\rb^c - \rb^g) \cdot \fb_g^c \leq 0$. The macroscopic object of interest is the stress tensor\cite{RothenburgSelvadurai:1981,*KruytRothenburg:1996}
\begin{equation}
\label{sigma}
\sigmab = -\frac{1}{V} \sum_{g\in G} \sum_{c\in C^g} (\rb^c - \rb^g) \fgc,
\end{equation}
where $V$ is the volume of the packing. In what follows we will decompose this into contributions from each grain $g$, $\sigmab^g \equiv 1/V^g \; \sum_{c\in C^g} (\rb^c - \rb^g) \fgc$, where $V^g$ is a volume associated to a grain, discussed below. 

If a continuum description exists, then mechanical equilibrium requires that the stress tensor satisfies \cite{ChaikinLubensky}
\begin{equation}
\label{Newton_cont}
0 = \nabla \cdot \sigmab, \;\;\; \sigmab = \sigmab^T.
\end{equation}
This is the macroscopic analog of \eqref{Newton}. In two dimensions (2D), in the absence of body forces, it implies that $\sigmab$ can be written as \cite{Muskhelishvili, *HenkesChakraborty:2009}
\begin{equation}
\label{Airy}
\sigmab = \nabla \times \nabla \times \psi,
\end{equation}
where $\psi$ is a scalar, known as the Airy stress function. In component form, 
\[
\sigmab = \begin{pmatrix} \;\; \p_{yy} \psi & -\p_{xy} \psi \\
                           -\p_{xy} \psi & \;\; \p_{xx} \psi \end{pmatrix}.
\]
In three dimensions (3D), \eqref{Airy} also holds, with $\psi$ replaced by a tensor $\psibt$, known as the Beltrami stress tensor. A $\sigmab$ of this form is identically divergence-free and symmetric, so $\psi$ automatically yields stress configurations in force and torque balance. 
The extension of \eqref{Airy} to granular materials was initiated by the seminal works of Satake \cite{Satake:1993,*Satake:1997,*Satake:2004} and Ball and Blumenfeld \cite{BallBlumenfeld:2002}. These authors found ``loop forces'' $\rhob$ such that $\sigmab = \nabla \times \rhob$, as well as partial analogs of $\psi$. However, they were not able to derive $\rhob$ from a discrete version of $\psi$ which gives identically torque-balanced configurations, so that the analogy with \eqref{Airy} is incomplete. 

In this paper we extend their results by deriving an exact, discrete analog of \eqref{Airy} for frictional sphere packings in two and three dimensions. 
We first present an \textit{intrinsic} formulation based solely on the \textit{topology} of the contact network. We then present a complementary \textit{extrinsic} formulation, by building the contact network into a triangulation of space. The added structure of the triangulation depends heavily on the \textit{geometry} of the contact network.

The new variables $\rhob$ and $\psi$ define changes of coordinates in phase space. In this paper, we consider the limit of large grain rigidity in which the deformation of grains can be ignored. 
This allows us to fix the grain positions and consider the Force Network Ensemble of force configurations on a fixed packing geometry \cite{Snoeijeretal:2004,*Tigheetal:2010}.  



The geometry consists of $N_{RG}$ real (force-bearing) grains as well as $N_{VG}$ virtual grains, or rattlers, which are trapped in the packing, but do not contribute to mechanical stability. The real grains touch at $N_{RC}$ force-bearing contacts \footnote{These include $N_{EC}$ external contacts at the boundary at which external forces are applied; we assume that each of the $N_{BG}$ boundary grains is subject to one external force.}, so phase space is spanned by the $N_{RC}$ contact forces and has dimension $d N_{RC}$. Force and torque balance restrict force configurations to a subset of phase space, of dimension
\begin{equation}
M_d = dN_{RC}-\frac{d(d+1)}{2} N_{RG} \label{Md}
\end{equation}
Writing $N_{RC}=\zbar N_{RG}/2$, we define the mean contact number $\zbar$. When $M_d=0$, the forces are uniquely determined by the grain positions; this is known as isostaticity \cite{Alexander:1998,EdwardsGrinev:1999} and occurs when $\zbar = d+1$. More generally we may also consider hyperstatic packings with $\zbar>d+1$ and $M_d>0$.

Since $M_d$ depends only on the geometry, it is a topological invariant under any coordinate change in force space. The intrinsic formulation uses the contact network, which assigns a vertex to each grain center and a link to each contact. Mechanical stability requires that real grains form closed loops $\ell$. This property implies a combinatorial identity between the number of independent loops $N_L$, real grains $N_{RG}$, and real contacts $N_{RC}$, which we can use to rewrite $M_d$. To see this identity, we consider an arbitrary grain $g$ and inductively build the entire packing, one loop at a time. Initially, we have one grain, no contacts, and no loops. We choose an arbitrary contact belonging to $g$ and trace out a shortest \footnote{'Shortest' means fewest number of contacts.} loop back to $g$, containing, say, $k$ new grains. In doing so we have added $k$ grains, $k+1$ contacts, and 1 loop, so $N_{RG}-N_{RC}+N_L$ is preserved at its original value, $1$. Continuing in this way, always beginning loops on existing grains, we eventually trace out the entire interior contact network, so that
\begin{equation}
\label{NL}
N_L=N_{RC}-N_{RG} + 1.
\end{equation}
This is a version of Euler's formula \footnote{Euler's formula states that $V-E+F=\chi$ for a graph with $V$ vertices, $E$ edges, and $F$ faces on a surface with Euler characteristic $\chi$. For a finite planar graph, we can embed the graph on the surface of a sphere, identifying the outside of the graph with a face that wraps around the sphere. This amounts to taking $\chi=1$ in this formula.} for the contact network. We call the set of independent loops chosen by this process a net. In 2D, the net is unique, but in higher dimensions, many nets are possible \cite{[{In graph theory, what we call loops are known as cycles, and $N_L$ is called the cyclomatic number, or nullity. If all grains have more than 2 real contacts, then nets are in one-to-one correspondence with spanning trees, which can be explicitly counted by Kirchoff's matrix-tree theorem. See }]Bollobas}. 

The relation \eqref{NL} allows us to write $M_d$ in two distinguished ways, by eliminating either $N_{RC}$, or $N_{RG}$. The first, 
\begin{equation}
M_d = d N_L - \frac{d(d-1)}{2} N_{RG} - d,\label{Md_rho}
\end{equation}
indicates that phase space is spanned by a vector field defined on the loops, providing $dN_L$ degrees of freedom, subject to torque balance, providing $\half d(d-1) N_{RG}$ constraints, and a single extra vector constraint. To exhibit this formulation explicitly, we assign to each loop a fixed orientation, and a pseudovector loop force $\rhob^\ell$. Setting $\rhob_g^\ell=+\rhob^\ell$ if the oriented loop $\ell$ enters the grain $g$, and $\rhob_g^\ell=-\rhob^\ell$ if it exits $g$, we now write each contact force as
\begin{equation}
\label{rhol}
\fgc = \sum_{\ell \in L^c} \rhob_g^\ell. 
\end{equation}
where $L^c$ is the set of loops adjacent to the contact $c$. This definition ensures that Newton's 3rd law is satisfied. Moreover, since each loop going through $g$ enters at precisely one contact, and exits at precisely one contact, when we sum the contact forces incident on a grain, each loop force appears in equal and opposite pairs, so that force balance is satisfied identically. 



In 2D, each contact is adjacent to two loops, and the loops can all be taken to have anticlockwise orientation. This makes \eqref{rhol} a simple difference of loop forces, so that adding a constant to each $\rhob^\ell$ leaves invariant the contact forces. In this case, there is a vector gauge freedom and so a vector constraint is needed to fix a gauge. However, in higher dimensions, the net is not unique, and there is no natural orientation of the loops. It is not clear in this case whether a gauge freedom exists, or whether the extra constraint is a net-dependent consistency requirement.

We may also write
\begin{equation}
M_d = \frac{d(d+1)}{2} N_L - \frac{d(d-1)}{2} N_{RC} - \frac{d(d+1)}{2}, \label{Md_loops}
\end{equation}
in which the number of grains no longer appears. 
This formulation is achieved by supplementing $\rhob^\ell$ with a field $\varphi^\ell$, a pseudoscalar in 2D and a pseudovector in 3D, defined so that
\begin{equation}
\label{varphil}
\rb^c \times \fgc = \sum_{\ell \in L^c} \left( \varphi_g^\ell + \rb^\ell \times \rhob_g^\ell \right).
\end{equation}
where $\rb^\ell$ is the position of $\ell$, discussed in the sequel. As with $\rhob_g^\ell$, $\varphi_g^\ell=+\varphi^\ell$ if the oriented loop $\ell$ enters the grain $g$, and $\varphi_g^\ell=-\varphi^\ell$ if it exits $g$. Summing the torques applied to a grain, it is easily seen that torque balance is satisfied identically.

To ensure that torques result from tangential contact forces, the torques inferred from $\rhob^\ell$ must equal those determined from $\varphi^\ell$, which requires
\begin{equation}
\label{varphiconsist}
0 = \sum_{\ell \in L^c} \left(\varphi_g^\ell + \left(\rb^\ell - \rb^c\right) \times \rhob_g^\ell \right). 
\end{equation}
These are the contact constraints appearing in \eqref{Md_loops}. As with the $\rhob$ formulation, the residual constraint may depend on the net. In two dimensions \eqref{rhol} and \eqref{varphil} reduce to Satake's formulation \cite{Satake:1993}.

The intrinsic formulation can be put into a differential form by defining topological divergence and curl operators. The former, defined for vector fields on the contacts, is $(\ds \bm{f})^g \equiv \sum_{c \in C^g} \fgc$, while the latter, defined for pseudovector and pseudoscalar fields on the loops, is $(\ds \rhob)_g^c \equiv \sum_{\ell \in L^c} \rhob^\ell_g$. These operators satisfy the identity $\ds\ds = 0$. Dropping subscripts and superscripts, we can write Newton's laws as $\ds \bm{f}=0$ and $\ds (\rb \times \bm{f})=0$, \eqref{rhol} as $\fb = \ds \rhob$, and \eqref{varphil} as $\rb \times \fb = \ds (\varphi + \rb \times \rhob)$.

Continuing in this fashion, one can define a complete topological calculus, with analogs of Stokes Theorem and Poincar{\'e}'s Lemma. There is a difficulty, however; vector fields, such as $\fb$, have no direct continuum meaning, since they change sign with the orientation of the contact. They are, properly, 1-forms. This difficulty may be overcome by resolving $\fb$ along a vector field; this is precisely what is accomplished by the stress tensor, \eqref{sigma}. This introduces geometry into the formulation. In 2D the loops tile the plane, so their geometry is simple, but in 3D, they may connect in a complex way. Moreover, in 3D nets are not uniquely defined, so given a contact $c$, there is no way of knowing how many loops are adjacent to $c$ without some knowledge of the whole net.

\begin{figure*}[ht!]
  \subfigure{\fbox{\includegraphics[viewport=80 70 400 300,width=0.46\textwidth,clip]{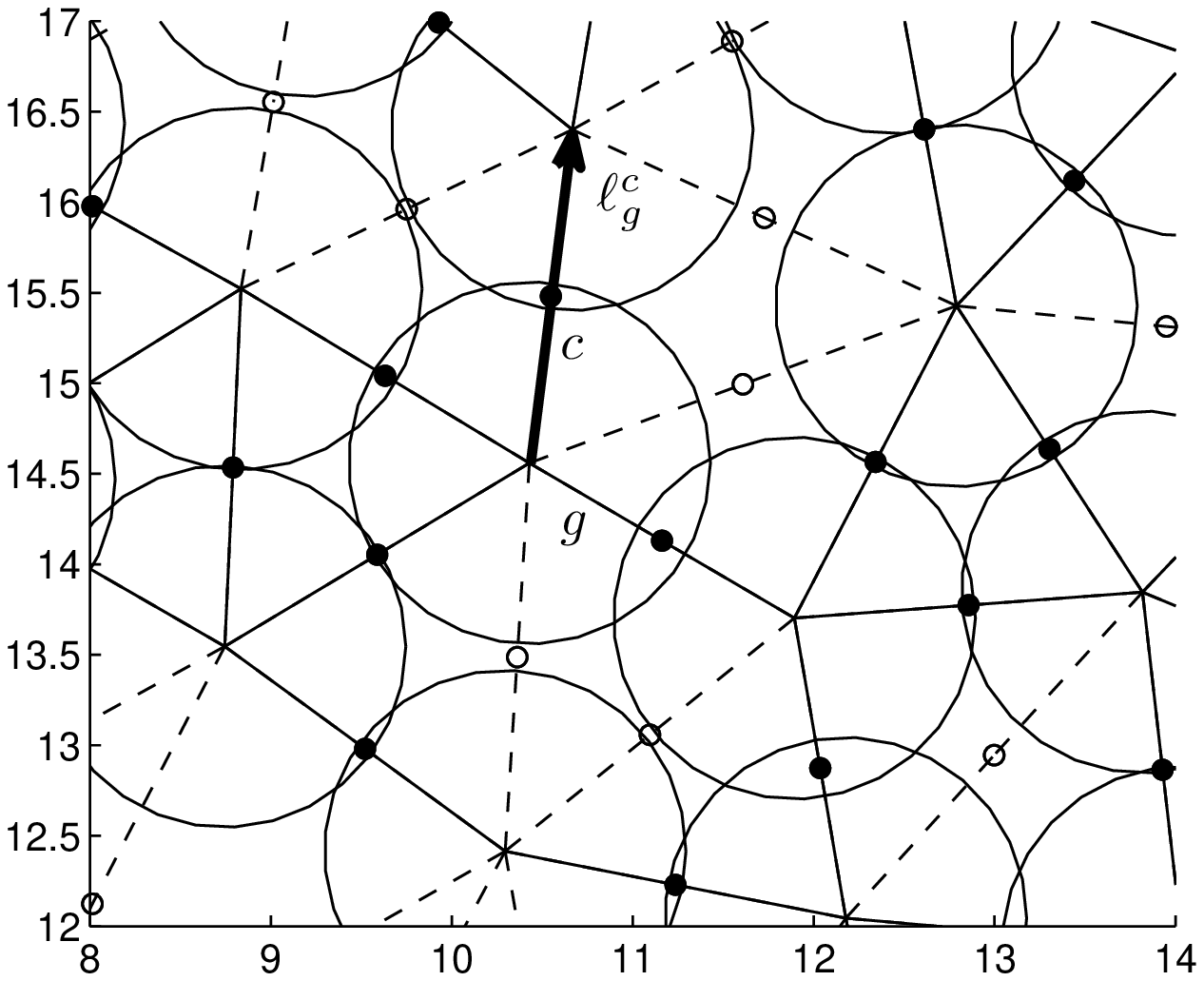}}}
  \subfigure{\fbox{\includegraphics[viewport=80 70 400 300,width=0.46\textwidth,clip]{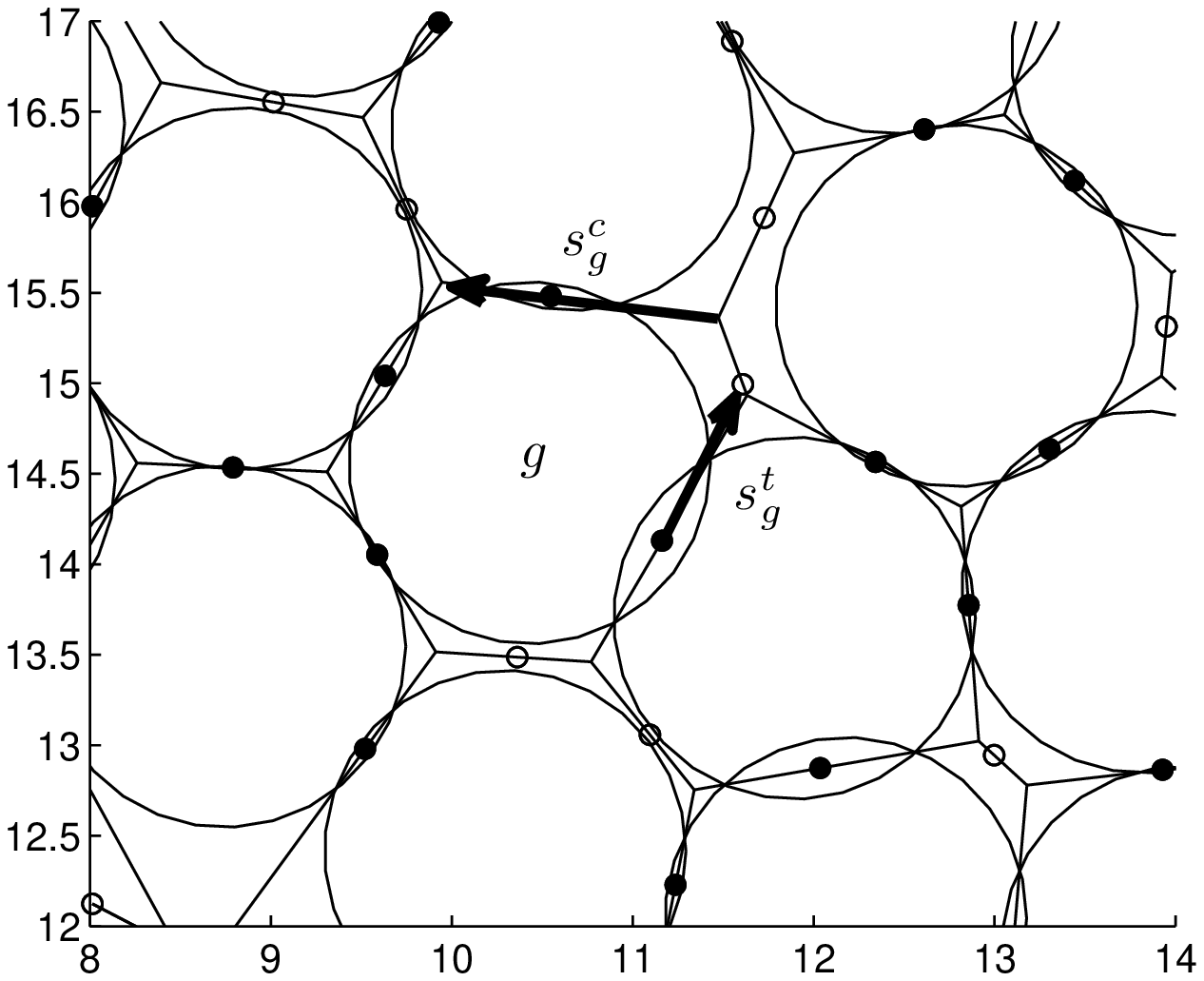}}}
  \caption[Delaunay triangulation and Voronoi tesselation]{\label{fig:geom2d} Delaunay triangulation (left) and Voronoi tesselation (right) in 2D. Real contacts are indicated by small dots, with their associated contact vectors solid. Virtual contacts are shown as unfilled circles, with their associated contact vectors dashed. Although in reality soft spheres deform on contact, for simplicity real contacts are shown with overlaps.}
\end{figure*}

Rather than proceeding with this admixture of topology and geometry, we 
therefore present a complementary extrinsic formulation based on a triangulation of space. By using a triangulation, the topology is locally regular and simply related to the geometry. Around any grain, we can determine the topology and geometry of the triangulation knowing only the positions of the grain's neighbours. In particular, we use the Delaunay triangulation and its dual, the Voronoi tessellation \footnote{Also known as the Dirichlet tessellation}, shown in Figure \ref{fig:geom2d}.

The Delaunay triangulation is formed by adding to the contact network links between adjacent grains which do not touch, so that space is partitioned into simplices: triangles in 2-space and tetrahedra in 3-space \footnote{This is done in 2-space (3-space) in such a way that the circumcircle (circumsphere) formed from each elementary triangle (tetrahedron) contains no other grains. Hence, the vertices of the triangles (tetrahedra) correspond to the nearest grain centers.}.  Physically, the new ``virtual'' contacts correspond to contacts that could be created under a small deformation \footnote{In fact, the triangulation provides a means of defining ``small deformation'' rigorously, as a deformation which preserves the topology of the triangulation. This partitions the phase space of grain positions into equivalence classes, the boundaries of which correspond to the swapping of virtual contacts across a loop.}. The Voronoi tesselation is the dual of the Delaunay triangulation; each Voronoi cell is centered on a grain and contains the points nearer to that grain than any other. We define the volume associated to a grain, $V^g$, as the volume of its Voronoi cell. 


Our general approach to derive an analog of \eqref{Airy} is to define an exact discrete calculus \cite{[{Elements of a discrete calculus have been reinvented many times in the literature. Our elementary approach emphasizes the connection with classical derivatives, whereas other works use the language of differential forms. Whereas we allow vector- and tensor-valued quantities on each graph element, others always have scalar-valued quantities, interpreting these as vectors or pseudovectors according to the graph element on which they are defined. See }]ItzyksonDrouffe:1991,*Schwalmetal:1999,*[{A more mathematical approach is presented in }]Desbrunetal:2005} on these graphs in which \eqref{Newton} appear as differential relations. By Poincar\'e's Lemma, $\nabla \cdot \bm{F} = 0 \Rightarrow \bm{F} = \nabla \times \bm{G}$, these suggest natural forms for $\rhob$ and $\psi$. Routine calculations then show that \eqref{Newton} holds identically in the new variables.

\subsection{Notation}
In this paper we deal with scalars, vectors, tensors, and their pseudo counterparts, defined on grains $g$, contacts $c$, triangles $t$, and tetrahedra $v$. When unambiguous, these are indicated by a single superscript, but it is often necessary to add subscripts to indicate orientation. These are explained as they appear.

The sets of all real grains, virtual grains, real contacts, virtual contacts, triangles, and tetrahedra in the packing are denoted by $RG$, $VG$, $RC$, $VC$, $T$, and $V$, respectively. Together, the real and virtual grains form the set of grains $G = RG \cup VG$, and similarly, the real and virtual contacts together form the set of contacts $C = RC \cup VC$. We also consider local sets, for example $C^g$ denotes the set of contacts surrounding a grain $g$, and likewise for other neighboring quantities. The sets of boundary contacts, triangles, and tetrahedra are denoted $BC, BT$, and $BV$, respectively.

Tensors are denoted with a hat. The identity tensor is written $\delb$. Outer (dyadic) products are simply denoted with a space. For vectors and 2-tensors, operators always act on adjacent indices, in the same order as they are normally written, and all contractions are indicated with a dot, the number of dots indicating the number of contracted indices; e.g., $\bm{\hat{A}}:\nabla (\bm{\hat{B}} \; (\nabla \cdot \bm{\hat{C}})) = \sum_{i,j,l} A_{ij} \p_i (B_{jk} (\p_l C_{lm}))$. We only need one 3-tensor, the 3D Levi-Civita symbol appearing in the 3D cross product. We take the free index to be the first, e.g., $(\bm{\hat{A}} \times \bm{\hat{B}})_{ijm} = \sum_{k,l} A_{ik} \varepsilon_{jkl} B_{lm}$. To conform to convention we make an exception to the above rules for the tensor curl in 3D, letting $(\nabla \times \bm{\hat{B}})_{ij} = \sum_{k} \varepsilon_{ikl} \p_k B_{jl}$.

In 2D the cross product is, e.g., $(\bm{\hat{A}} \times \bm{\hat{B}})_{il} = \sum_{j,k} A_{ij} \varepsilon_{jk} B_{kl}$, with $\epsb$ the 2D Levi-Civita symbol, $\varepsilon_{12} = -\varepsilon_{21} = 1$, $\varepsilon_{11} = \varepsilon_{22} = 0$. The 2D curl is $(\nabla \times \bm{u})_{ij} = \sum_{k} \varepsilon_{ik} \p_k u_{j}$. In both 2D and 3D, the condition that a 2-tensor $\bm{\hat{A}}$ be symmetric can be written
\begin{equation}
\label{symmetric}
\bm{\hat{A}} = \bm{\hat{A}}^T \iff \epsb:\bm{\hat{A}}=0.
\end{equation}

Contours are always oriented anticlockwise, and we use $+/-$ to indicate geometric elements anticlockwise/clockwise from given elements, around grains, triangles, etc. For example, in 2D, $c^+=c^+(g,t)$ is the contact anticlockwise from triangle $t$ when looking from grain $g$. 
\section{\label{sect:2D}2D}
In the plane, contacts correspond to edges of Voronoi cells; we assign vectors $\bm{s}_g^{c}$ along these edges, circulating anticlockwise around grains (Figure \ref{fig:geom2d}). We also assign vectors $\lgc$ pointing from the center of $g$ to the center of its neighbour at $c$. These allow a natural definition for the area of a contact, $A^c = \half \; \lb^c \times \bm{s}^{c}$. For a tensor field defined on the contacts, we define its divergence by analogy with a tensor version of Gauss' Theorem, defining the resulting line integral with the $\bm{s}$ vectors:
\begin{equation*}
\int_g dA \; \nabla \cdot \sigmab \equiv A^g \left(\nabla \cdot \sigmab \right)^g \equiv \mm \sum_{c \in C^g} \bm{s}_g^{c} \times \sigmab^c \equiv \mm \oint_{\p g} d\bm{r} \times \sigmab.
\end{equation*}
We emphasize that the underlying fields and definitions are discrete. Because the discrete fields satisfy a similar algebra as in the continuum, the continuum notation is useful for intuition and calculations. 

Rewriting \eqref{sigma} as a sum over contacts, we have $A^c \sigmab^c = -\lb^c \fb^c$. From this definition we see that $\bm{s}_g^{c} \times \sigmab^c = 2 \fb^c$, so $\left(\nabla \cdot \sigmab \right)^g = 0$ is equivalent to force balance. By inspection of \eqref{sigma}, and using \eqref{symmetric}, we see that $\sigmab^g = (\sigmab^g)^T$ is equivalent to torque balance, and hence we reproduce the expected continuum equations \eqref{Newton_cont} in the discrete calculus, at the grain scale.

We will define the pseudovector $\rhob$ on the triangles, which correspond to vertices of Voronoi cells; we assign vectors $\bm{s}_g^{t}$ circulating anticlockwise around grains, connecting the Voronoi vertices. Using a form of Green's Theorem, for a vector field defined on the triangles, we may define its curl as
\begin{equation}
\label{curlrho}
\int_g dA \; \nabla \times \rhob \equiv A^g \left(\nabla \times \rhob \right)^g \equiv - \sum_{t \in T^g} \bm{s}_g^{t} \; \rhob^t \equiv - \oint_{\p g} d\rb \; \rhob.
\end{equation}
Noting that $2 \bm{s}_g^{t} = \lgcp - \lgcm$, we see that $\sigmab^g = (\nabla \times \rhob)^g$ if
\begin{equation}
\label{rhot}
\fb_g^c = \rhob^{t^-} - \rhob^{t^+},
\end{equation}
where $t^+=t^+(g,c)$ is the triangle anticlockwise from contact $c$. Summing the contact forces incident on a grain, all loop forces appear in equal and opposite pairs, so force balance is identically satisfied. 

By working with the Delaunay triangulation, we treat real and virtual contacts on the same footing. Therefore, equation \eqref{rhot} applies to both real and virtual contacts, and hence generically leads to virtual contact forces. To obtain physical force configurations, we must explicitly impose that these vanish: if $c \in VC^g, 0 = \fgc = \rhob^{t^-} - \rhob^{t^+}$. Since virtual contacts only occur in the interior of loops, this shows that $\rhob^t$ must be constant on loops; this also shows that we recover the formulation of Satake \cite{Satake:1993,*Satake:1997,*Satake:2004} and Ball and Blumenfeld \cite{BallBlumenfeld:2002} if we sum over the virtual contacts. 

An immediate consequence of $\sigmab^g = (\nabla \times \rhob)^g$ is that the stress tensor for the packing can be written as a boundary sum \cite{KruytRothenburg:1996}
\begin{equation*}
\int_{\Omega} dA \; \sigmab \equiv A \sigmab = - \sum_{t \in B T} \bm{s}^{t} \rhob^t \equiv - \oint_{\p \Omega} d\rb \; \rhob,
\end{equation*}
where the $\bm{s}$'s connect boundary contacts in an anticlockwise manner. 

We may check that $M_2$ is conserved under this coordinate change by counting triangles. Applying Euler's formula to a single loop $\ell$, $N_{VG}^\ell - N_{VC}^\ell + N_T^\ell=1$. Adding this relation over the whole packing, we find $N_{VG} - N_{VC} + N_T = N_L$, so that, using \eqref{Md}, $M_{2} = 2N_T - (2N_{VC} + N_{RG} - 2N_{VG} +2)$. This indicates that the loop forces are constrained by the virtual contact constraints and by torque balance, and that $2N_{VG}$ constraints become redundant. This redundancy is a consequence of the definition of loop forces. Indeed, since loop forces automatically yield grains in force balance, for a virtual grain with $z$ virtual contacts, it is sufficient to impose $z-1$ virtual contact contraints to guarantee that all $z$ virtual contact forces vanish. Finally, the loop forces have a gauge freedom $\rhob \rightarrow \rhob + \Delta \rhob$, with $\Delta \rhob$ any constant, as is clear from their definition.

The torque balance constraint can be rewritten in terms of $\rhob$ by first defining
\begin{equation*}
\int_g dA \; \nabla \cdot \rhob \equiv A^g \left(\nabla \cdot \rhob \right)^g \equiv - \sum_{t \in T^g} \bm{s}_g^{t} \times \rhob^t \equiv - \oint_{\p g} d\rb \times \rhob.
\end{equation*}
Comparing this equation with \eqref{curlrho}, and using $\sigmab^g = (\nabla \times \rhob)^g$, we see by inspection that $\sigmab^g = (\sigmab^g)^T$ is equivalent to  $(\nabla \cdot \rhob)^g = 0$. This motivates a search for $\psi^c$ such that $\rhob^t = \nabla \times \psi^c$. We define
\begin{equation*}
\int_t dA \; \nabla \times \psi \equiv A^t \left(\nabla \times \psi \right)^t \equiv - \sum_{c \in C^t} \lb_t^c \; \psi^c \equiv - \oint_{\p t} d\rb \; \psi.
\end{equation*}
It is not obvious from this definition, but as shown in Appendix A \ref{app:torque}, this yields force configurations that identically satisfy torque balance. Since each triangle is surrounded by three contacts, and contacts interior to a loop are shared by two triangles, $3N_T^\ell = 2N_{VC}^\ell + N_{RC}^\ell$. This yields the global counting relation $3N_T = 2N_C + N_{B C}$, in which $B C$ is the set of triangle edges that circulate around the boundary. We can therefore write $M_{2} = (N_C +N_{B C}) - (2N_{VC} - 3N_{VG} +3)$. The stress function $\psi$ is defined on the real and virtual contacts, and constrained on the virtual contacts by the requirement of no virtual contact force. Because rattlers automatically satisfy force and torque balance, $3N_{VG}$ constraints are redundant. Finally, $\psi$ has a three dimensional gauge freedom $\psi \rightarrow \psi + \Delta \psi$, with $(\Delta \psi)^c = \bm{A} \cdot \bm{r}^c + B$ any linear function of position, which follows from the computation $\nabla \times (\Delta \psi)^c = \bm{\epsb} \cdot \bm{A}$.


We have therefore succeeded in writing $\sigmab^g = \nabla \times \rhob^t = \nabla \times \nabla \times \psi^c$. It is possible to invert this transformation, up to gauge freedom. For a contour of adjacent triangles $(t_0,t_1,\ldots,t_n)$, separated by contacts $(c_1,c_2,\ldots,c_n)$, we find
\begin{equation}
\label{rholine}
\rhob^{t_n} - \rhob^{t_0}  = \sum_{i=1}^{n} \fb^{c_i} =  \half \sum_{i=1}^{n} \bm{s}^{c_i} \times \sigmab^{c_i} \equiv \half \int_{t_0}^{t_n} d\rb \times \sigmab,
\end{equation}
where the forces are exerted from the left side of the contour to the right side. Similarly,
\begin{equation}
\label{psiline}
\psi^{c_n} - \psi^{c_0} = - \sum_{i=0}^{n-1} \bm{s}^{t_i} \times \rhob^{t_i} \equiv - \int_{c_0}^{c_n} d\rb \times \rhob.
\end{equation}
Force and torque balance ensure that the line integrals in \eqref{rholine} and \eqref{psiline}, respectively, are path-independent.

In the 2D discrete calculus, we therefore have the theorems
\begin{equation}
\label{thm1}
\nabla \cdot \sigmab^c = 0 \;\; \iff \;\; \sigmab^g = \nabla \times \rhob^t
\end{equation}
and
\begin{equation}
\label{thm2}
\nabla \cdot \rhob^t = 0 \;\; \iff \;\; \rhob^t = \nabla \times \psi^c.
\end{equation}

Equation \eqref{psiline} allows us to relate $\psi^c$ to $\varphi^\ell$ in the intrinsic formulation. We first allow $\psi$ to take two values on each real contact, one per adjacent loop $\pm\ell$. We define a mean $\psi$ within a loop by $z^\ell \bar{\psi}^\ell \equiv \sum_{c \in RC^\ell} \psi_\ell^c$, where $z^\ell$ is the number of contacts around the loop. Then \eqref{psiline} implies $\psi_\ell^c = \bar{\psi}^\ell + (\bm{r}^\ell - \bm{r}^c) \times \rhob^\ell$, where $z^\ell \bm{r}^\ell \equiv \sum_{c \in RC^\ell} \bm{r}^c$. Comparing with \eqref{varphiconsist} we see that $\varphi^\ell = \bar{\psi}^\ell$ and the consistency constraints in the intrinsic formulation are equivalent to $\psi_{+\ell}^c = \psi_{-\ell}^c$, a type of Newton's 3rd law. 

\begin{figure}[ht!]
  \fbox{\put(60,48){$g$}\put(60,86){$c$}\put(44,71){$\bm{s}^t_{gc}$}\put(60,121){$\bm{\ell}_g^c$}
\includegraphics[width=0.4\textwidth,trim=70 70 130 140,clip]{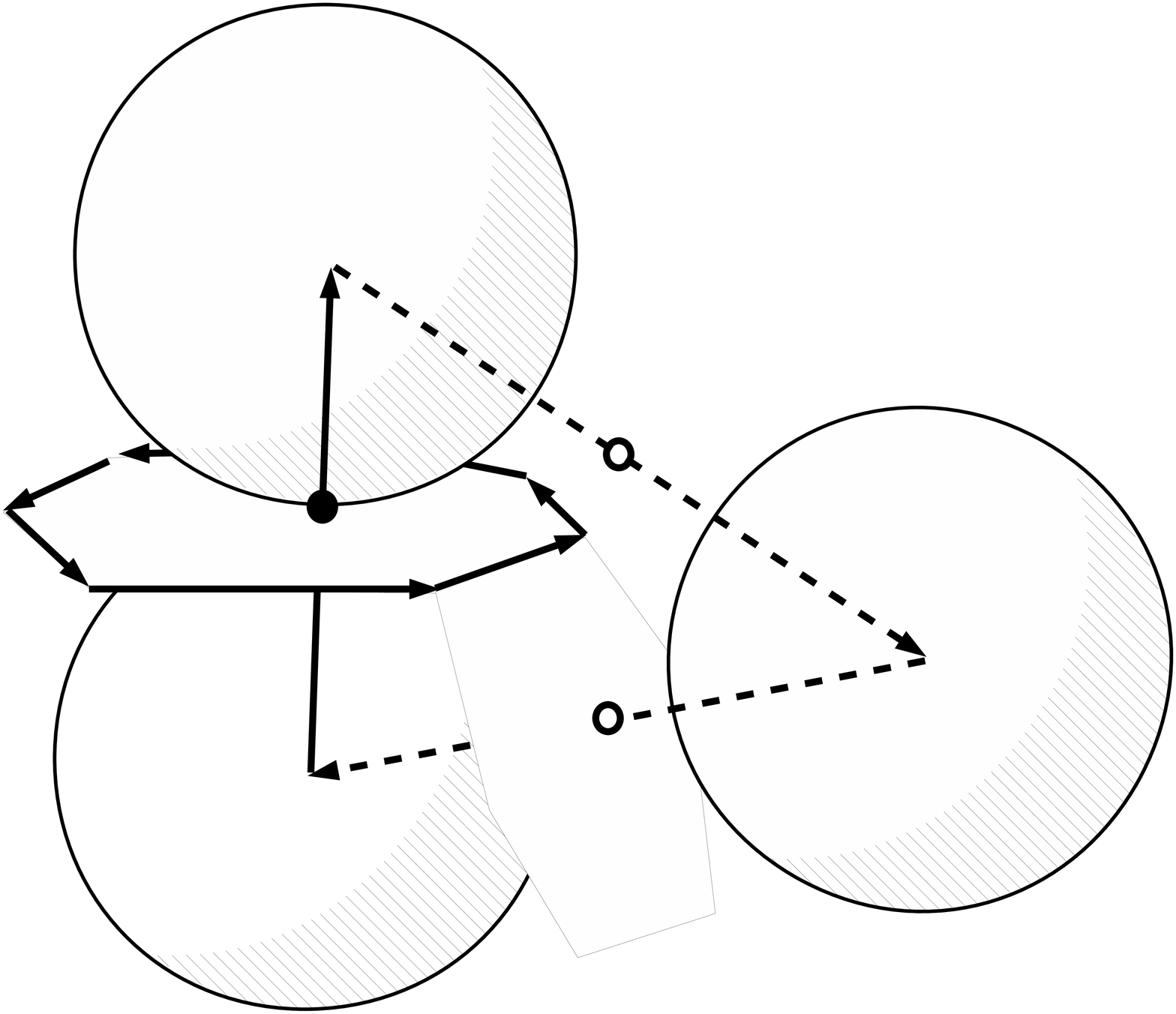}}
  \caption[Subset of Delaunay triangulation and Voronoi tesselation]{\label{fig:geom3d} Subset of Delaunay triangulation and Voronoi tesselation in 3D. Real contacts are indicated by small dots, with their associated contact vectors solid. Virtual contacts are shown as unfilled circles, with their associated contact vectors dashed.}
\end{figure}
\section{\label{sect:3D}3D}

In three dimensions, we can proceed analogously. The Voronoi tesselation assigns to each grain a convex polyhedron (Figure \ref{fig:geom3d}). Each face corresponds to a contact, each edge corresponds to a Delaunay triangle, and each vertex corresponds to a Delaunay tetrahedron. As before, we can define a discrete divergence with Gauss' Theorem:
\begin{equation*}
\int_g dV \; \nabla \cdot \sigmab \equiv V^g \left(\nabla \cdot \sigmab \right)^g \equiv \sum_{c \in C^g} A^c \ltildeb_g^c \cdot \sigmab^c \equiv \int_{\p g} d\bm{S} \cdot \sigmab,
\end{equation*}
where $\ltildeb = \lb/|\lb|$ and the natural area for a contact, $A^c$, is the area of the contact face. Rewriting \eqref{sigma} as a sum over contacts, we have $V^c \sigmab^c = -\lb^c \fb^c$. The natural choice of $V^c$ is the volume of the bipyramid whose vertices are the neighbouring grain centers and whose base is the contact face. Using $V^c = \mbox{$\frac{1}{3}$} A^c |\lb^c|$, we see that $A^c  \ltildeb_g^c \cdot \sigmab^c = -3 \ltildeb_g^c \cdot \ltildeb^c \fb^c = -3 \fb_g^c $, so again $\left(\nabla \cdot \sigmab \right)^g = 0$ is equivalent to force balance. Again $\sigmab^g = (\sigmab^g)^T$ is equivalent to torque balance, and hence we reproduce the expected continuum equations \eqref{Newton_cont} in the discrete calculus, at the grain scale. 

We now seek to define $\rhobt$ so that $\sigmab = \nabla \times \rhobt$. By Gauss' Theorem
\begin{equation*}
\int_g dV \; \nabla \times \rhobt \equiv V^g \left(\nabla \times \rhobt \right)^g \equiv \!\sum_{t \in T^g} d\bm{S}_g^t \times (\rhobt^t)^T \equiv \!\int_{\p g} d\bm{S} \times \rhobt^T. 
\end{equation*}
The natural way to define $d\bm{S}_g^t$ is to decompose the contact faces on either side of $t$ into triangles of area $A_{c^{\pm}}^t$ and set $d\bm{S}_g^t= A_{c^+}^t \ltildeb_g^{c^+} + A_{c^-}^t \ltildeb_g^{c^-} \equiv \sum_{c \in C_g^t} A_{c}^t \ltildeb_g^{c}$. It is easily seen from Figure \ref{fig:geom3d} that $A_{c}^t \ltildeb_g^{c} = \half (\rb^t-\rb^c) \times \bm{s}_{gc}^t$, where $\rb^t$ is the intersection of the triangle $t$ with the corresponding edge of the Voronoi cell of $g$, and $\bm{s}_{gc}^{t}$ circulates anticlockwise around contacts. We can now rewrite $\left(\nabla \times \rhobt \right)^g$ as
\begin{align*}
V^g \left(\nabla \times \rhobt \right)^g & = -\half \sum_{t \in T^g} \sum_{c \in C_g^t} \left((\rb^c-\rb^t) \times \bm{s}_{gc}^t \right) \times (\rhobt^t)^T \\
& = -\half \sum_{t \in T^g} \sum_{c \in C_g^t} \left((\rb^c-\rb^g) \times \bm{s}_{gc}^t \right) \times (\rhobt^t)^T \\
& = -\quarter \sum_{c \in C^g} \sum_{t \in T^c} \left(\lb_g^c \times \bm{s}_{gc}^t \right) \times (\rhobt^t)^T,
\end{align*}
where we have used the identity $\sum_{c \in C^t_g} \sgct = 0$ to exchange $\rb^t$ with $\rb^g$. Using \eqref{sigma}, we now see that $\sigmab^g = \nabla \times \rhobt^t$ if $\lgc \fgc = \half \sum_{t \in T^c} \left(\lb_g^c \times \bm{s}_{gc}^t \right) \times (\rhobt^t)^T$. This is equivalent to  
\begin{equation}
\label{loopforce3d1}
\bm{f}_g^c = - \half \sum_{t \in T^c} \rhobt^t  \cdot \bm{s}_{gc}^t \equiv - \half \oint_{(\p c)_g} \rhobt \cdot d\rb,
\end{equation}
and $0 = \sum_{t \in T^c} \bm{s}_{gc}^t \rhobt^t \cdot \lb_g^c$. The latter equation is satisfied identically if we let $\rhobt^t = -\ffrac{2}{|\bm{s}^t|^2} \rhob_{gc}^t \; \bm{s}_{gc}^t$. This allows \eqref{loopforce3d1} to be rewritten $\bm{f}_g^c = \sum_{t \in T^c} \rhob_{gc}^t$, which shows explicitly that the contact forces depend only on a vectorial quantity, albeit one with an intrinsic orientation. It also shows reduction to the form of the intrinsic formulation.

Summing the contact forces incident on a grain, all loop forces appear in equal and opposite pairs, so force balance is identically satisfied. As in 2D, we must explicitly require that there be no virtual contact forces. A new feature in 3D is that $\rhob_{gc}^t$ has a nontrivial gauge freedom $\rhob_{gc}^t \rightarrow \rhob_{gc}^t + (\Delta \rhob)_{gc}^t$ with $(\Delta \rhob)_{gc}^t = \bm{B}^{v^+} - \bm{B}^{v^-}$, for any vector field $\bm{B}^{v}$ defined on the Delaunay tetrahedra, which are dual to Voronoi vertices.

Again, $\sigmab^g = (\nabla \times \rhob)^g$ implies that the stress tensor for the packing can be written as a boundary sum \cite{KruytRothenburg:1996}
\begin{equation*}
\int_{\Omega} dV \; \sigmab \equiv V \sigmab = \sum_{c \in B C} A^c \bm{l}^{c} \times (\rhob^t)^T \equiv \oint_{\p \Omega} d\bm{S} \times \rhob^T,
\end{equation*}
where the $\bm{l}$'s are oriented outwards.


We now seek $\psibt^c$ such that $\rhobt^t = \nabla \times \psibt^c$. 
Using Stokes' Theorem, we set
\begin{equation*}
\int_t (\nabla \times \psibt) \cdot d\bm{S} \equiv A^t \left(\nabla \times \psibt \right)^t \cdot \stildeb^t \equiv \sum_{c \in C^t} \psibt^c \cdot \lb^c \equiv \int_{\p t} \psibt \cdot d\bm{r},
\end{equation*}
where $\stildeb = \bm{s}/|\bm{s}|$ and the contour is oriented anticlockwise around $\stildeb^t$. A natural choice to guarantee $\rhobt^t = \nabla \times \psibt^c$ is to set $\psibt^c = \quarter \ltildeb^c \ltildeb^c \psi^c$; then we have $A^t \rhob_{gc}^t = -\eighth |\bm{s}^t| \sum_{c \in C^t} \lb_{gt}^c \psi^c$, in close analogy with the 2D case. The computation that this leads to torque-balanced force configurations is almost identical to the 2D case, and for the same reasons, works only for identical spheres. $\psi$ has a complex gauge freedom, discussed in Appendix B \ref{app:dof}.


\section{Discussion and Extensions}

\subsection{Physical Interpretation}
As gauge variables, $\rhob$ and $\psi$ do not admit a unique physical interpretation. However, in 2D, the inversion formulae suggest a macroscopic interpretation of potential differences. In particular, equation \eqref{rholine} indicates that a difference of loop forces gives the flux of stress across any contour connecting two triangles. Similarly, equation \eqref{psiline} indicates that a difference of stress function gives the flux of loop force across any contour connecting two contacts. A linear change in stress function gives a constant flux of loop force, and hence no stress. Therefore, stresses correspond to curvature of the stress function. We prove in Appendix C that in 2D, in regions where macroscopic normal stresses are repulsive, the stress function is convex. 

By considering how a single contact force is written in terms of $\psi$, it is also possible to obtain a microscopic interpretation. In 2D, an arbitrary contact force $\fgc$ can be decomposed as
\begin{equation}
\fgc = \left( \frac{1}{A^{\ell^+}} + \frac{1}{A^{\ell^-}} \right) \lgc \psi^c + \sum_{c' \in C^{c}} \frac{1}{A^{\ell(c')}} \lb^{c'} \psi^{c'},
\end{equation}
where $C^c$ is the set of contacts surrounding the loops adjacent to $c$, and the $\lb^{c'}$ are oriented in the same direction as $\lgc$. Considering rigid grains so that the geometry is fixed, we see from this expression that an increase in $\psi^c$ directly increases the normal force at $c$. It also propagates to neighboring contacts, adding to their contact forces a component in the direction of $\lgc$. If the neighboring loops contain virtual contacts, the disturbance further propagates to the surrounding contacts in such a way that force and torque balance are preserved. We can therefore associate a change in $\psi^c$ with a change in the normal force at $c$, as well as the necessary changes in normal and tangential forces at neighbours of $c$ so as to preserve force- and torque-balance.

\begin{figure}[ht!]
  \centering
 \fbox{
\begin{lpic}[nofigure,l(-2mm),r(1.5mm)]{"ReciprocalDiagram"(3.2in)}
\includegraphics[viewport=90 100 500 400,width=0.46\textwidth,clip]{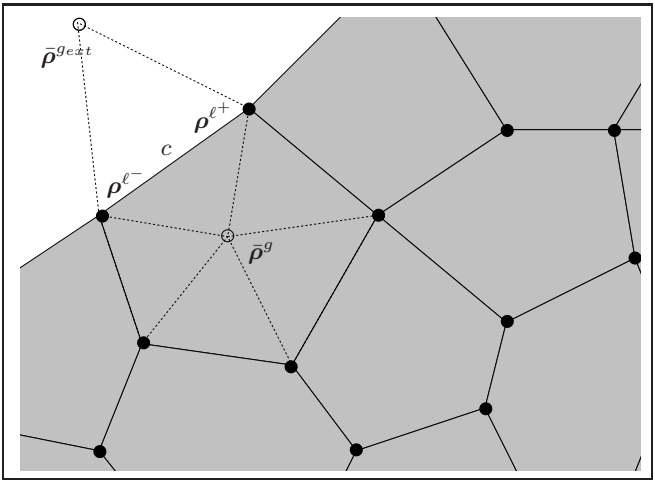}
   \lbl[bl]{-220,145,0;$c$}
   \lbl[bl]{-180,95,0;$\bar{\rhob}^g$}
   \lbl[bl]{-275,185,0;$\bar{\rhob}^{g_{ext}}$}
   \lbl[bl]{-245,125,0;$\rhob^{\ell^-}$}
   \lbl[bl]{-205,155,0;$\rhob^{\ell^+}$}
\end{lpic}

 }
  \caption[Maxwell-Cremona Diagram]{\label{fig:maxwell} Subset of a Maxwell-Cremona reciprocal diagram. Each polygon corresponds to a single grain. Also shown is the subdivision into triangles for one particular grain, and a triangle corresponding to an external contact.}
\end{figure}

In 2D, one can also obtain some geometrical intuition for these quantities by plotting the loop forces as points in force space, and drawing lines connecting the loop forces of adjacent loops, which are the contact forces. The Maxwell-Cremona \textit{reciprocal diagram} thus obtained \cite{Maxwell:1864,*Tigheetal:2008} is a tiling of polygons, each tile corresponding to a single grain (Figure \ref{fig:maxwell}). The total area of the tiling depends only on the boundary loop forces, and hence only on the boundary contact forces. It can be estimated for large packings, as follows. 

We define nominal tile centers for each grain, $\bar{\rhob}^g$, and divide each tile into triangles, one for each contact. The signed area of each triangle is $a^c_g =  \ffrac{1}{2} (\rhob^{\ell^-} - \bar{\rhob}^g) \times (\rhob^{\ell^+} - \bar{\rhob}^g)$. For interior contacts, adding the contributions from each adjacent grain, we have
\eq{
a^c \equiv a^c_{g^+} + a^c_{g^-} = \ffrac{1}{2}  \fb^c_{g^-} \times (\bar{\rhob}^{g^+} - \bar{\rhob}^{g^-}). 
}
The total area of the tiling is
\eq{ \label{area}
\mathcal{A} = \sum_{c \in RC} a^c - \sum_{c \in EC} a^c_{g_{ext}},
}
where the second term sums over the $N_{EC}$ external contacts at the boundary, to correct for overcounting in the first sum. It is clear from Figure \ref{fig:maxwell} that $\mathcal{A}$ is independent of the choice of $\bar{\rhob}^g$, but the relative size of the terms in \eqref{area} \textit{is} sensitive to this choice. For example, the trivial choice $\bar{\rhob}^g=0$ makes all $a^c=0$, and the entire contribution comes from the boundary term.

This work has shown that $\sigmab = \nabla \times \rhob$, which implies that the mean-field variation of $\rhob$ is given by $\bar{\rhob} (\rb)= \sigmab \times \rb$ in the continuum, up to an irrelevant constant of integration. If we choose $\bar{\rhob}^g = \sigmab \times \rb^g$, then, on average, each $\bar{\rhob}^g$ is in the middle of the tile corresponding to $g$, as shown in Figure \ref{fig:maxwell}. The sums in \eqref{area} therefore scale with the number of terms, i.e, as $N_{RC} \sim N$, and $N_{EC} \sim \sqrt{N}$, respectively, and hence for large enough packings, the boundary term is negligible. In fact, with this choice of $\bar{\rhob}^g$, the interior term is exactly 
\eq{
\sum_{c \in RC} a^c & = \ffrac{1}{2} (\epsb \cdot \sigmab \cdot \epsb) : \sum_{c \in RC} \fb^c_{g} \lb^c_g \\
& = - \ffrac{1}{2} (\epsb \cdot \sigmab \cdot \epsb) : A \sigmab^T \\
& = A \det(\sigmab),
}
so that $\mathcal{A} = A \det(\sigmab) + O(\sqrt{N})$. For periodic packings, the boundary term vanishes identically \cite{TigheVlugt:2010}. 

The above choice of $\bar{\rhob}^g$ does not guarantee that each $a^c$ is positive, which might be desirable in applications \cite{Tigheetal:2008}. If we choose, instead, $\bar{\rhob}^g = \det(\sigmab)/P \; \epsb \cdot \rb^g$, where $P = 1/2 \; \tr(\sigmab)$ is the pressure, then we find $a^c = -\det(\sigmab)/(2P) \; \fb^c_{g^-} \cdot \lb^c_{g^-}$. In this representation, \textit{repulsion of the contact forces is equivalent to positivity of the areas}. Moreover, $\sum_{c \in RC} a^c = A \det(\sigmab)$ as before, so the boundary term must again be negligible for large enough packings.

Since the tiling area depends only on the boundary forces, it is invariant under force-balance preserving changes of interior forces. In our formulation, localized force rearrangements can be generated by changing the stress function $\psi^c$ at a particular contact, which changes the loop forces at adjacent loops. This construction can therefore be used to explore the Force Network Ensemble, and serves the same purpose as the ``wheel moves" of Tighe and collaborators \cite{Tigheetal:2005,*TigheVlugt:2010,*TigheVlugt:2011}.

Attempts to extend the reciprocal diagram construction to 3D have not yet been successful \cite{TigheVlugt:2010}. In fact, since the loop forces are proper pseudovectors in 3D, they cannot simply be plotted as points in space and therefore cannot define the vertices of any polyhedra. Hence, if an analog of the reciprocal diagram exists, it is not obviously related to the loop forces defined in this work.

\subsection{Polydisperse Packings}
For clarity we have considered only packings of identical d-spheres, but many of our results extend to polydisperse packings. Indeed, because it is topological in nature, the intrinsic formulation \eqref{rhol}, \eqref{varphil}, and \eqref{varphiconsist} holds for packings of arbitrary convex grains, without modification. 

To extend the extrinsic formulation to polydisperse spheres, it is most convenient to use the radical Voronoi tesselation \cite{GellatlyFinney:1982}, which ensures that the edges of Voronoi cells pass through contacts. Discrete derivatives can be defined exactly as in this paper, and the definition of $\rhob$ is easily made, so that $\sigmab = \nabla \times \rhob$. However, $\psi$ as defined in this paper does \textit{not} describe identically torque-balanced configurations. The extension of $\psi$ to the polydisperse case will be discussed in future work.

\subsection{Force Law}

For simplicity, the discussion above was limited to hard disks and spheres, interacting without cohesion. In fact, these assumptions are inessential. 

We considered hard disks and spheres, and framed this paper as a discussion of force configurations on a fixed geometry, to show the power of degree-of-freedom counting. However, the definitions of $\rhob$ and $\varphi$ hold for disks and spheres of arbitrary softness. Provided contacts appear at the midpoint of contact vectors, the definition of $\psi$ also holds.

We also considered purely repulsive packings. In a repulsive packing, all real grains form closed loops. When grains admit cohesive forces, in addition to the loop-forming grains which bear the external load, there may be tree-like subsets of grains which extend into loops. Starting from the leaves of the tree, it is easy to see that each contact force on each of these grains must vanish. Therefore, provided the grains are treated as virtual, all of the results of this paper hold without modification. The physical novelty of cohesion is that the vanishing contact forces may be composed of an adhesive component, for example due to hydrodynamic forces, as well as an elastic component. This allows the grains to remain in place under an infinitesmal disturbance, unlike the rattlers.

More generally, this discussion shows that our results have little to do with the form of the force law and the repulsive constraints at the contacts, although the latter are an essential feature of granular physics \cite{Moukarzel:1998}. Rather, the key physical requirement is that grains interact locally, so that a well-defined contact network exists. Our results then follow from the general form of Newton's laws, \eqref{Newton}, acting on this network. Our results may therefore be useful in other problems involving local balance constraints, e.g. rigidity percolation \cite{JacobsThorpe:1996} and metabolic networks \cite{Segreetal:2002,*Almaasetal:2004}.

\section{Conclusion}
We have shown that Newton's laws \eqref{Newton} can be interpreted as differential relations in a discrete calculus. In an intrinsic topological formulation, these take the form
\eq{ \label{int1}
\ds \bm{f}=0, \qquad \ds (\rb \times \bm{f})=0.
}
These fields can be written in terms of gauge fields $\rhob$ and $\varphi$ as 
\eq{ \label{int2}
\fb = \ds \rhob, \qquad \rb \times \fb = \ds (\varphi + \rb \times \rhob),
}
which are required to satisfy, for consistency,
\eq{ \label{int3}
\ds (\varphi + \rb \times \rhob) = \rb \times \ds \rhob.
}
As discussed above, the stress tensor does not appear in this formulation, so the relation \eqref{Airy} is absent. Equations \eqref{int1}, \eqref{int2} and \eqref{int3} are exact, and hold for polydisperse, cohesive or non-cohesive, soft d-sphere packings. In this paper we considered $d=2$ (disks) and $d=3$ (spheres), but their extension to higher dimensions is straightforward.

To make contact with continuum equations, we also constructed an extrinsic formulation using the Delaunay triangulation and Voronoi tesselation. Using the triangulation, we constructed a discrete calculus in which Newton's laws \eqref{Newton} reproduce their continuum equivalent \eqref{Newton_cont} at the scale of a \textit{single grain}, i.e.,
\eq{ \label{ext1}
\left(\nabla \cdot \sigmab \right)^g = 0, \qquad \sigmab^g = (\sigmab^g)^T.
}
In 2D, we introduced gauge fields $\rhob$ and $\psi$, defined so that
\eq{ \label{gauge}
\sigmab^g = (\nabla \times \rhob)^g, \qquad \rhob^t = (\nabla \times \psi)^t.
}
The same relations hold in 3D with $\rhob \to \rhobt$ and $\psi \to \psibt$. To obtain physical force configurations, these must be subject to the geometry-dependent virtual contact constraint $\sigmab^c|_{c \in VC} = 0$. Equations \eqref{ext1}, and \eqref{gauge} are exact, and hold for monodisperse, cohesive or non-cohesive, packings of soft disks or spheres.

Together, \eqref{gauge} imply that $\sigmab^g = (\nabla \times \nabla \times \psi)^g$, which is an exact, discrete representation of \eqref{Airy}. On a homogeneous packing geometry, in the continuum limit, $\sigmab$ will satisfy $\sigmab = \nabla \times \nabla \times \psi$, with $\psi \to \psibt$ in 3D. From this expression we see that the pressure
\eq{ \label{pressure}
P \equiv \frac{1}{d} \; \tr(\sigmab) =  \begin{cases}
  \ffrac{1}{2} \nabla^2 \psi & \text{in 2D} \\
  \ffrac{1}{3} \nabla^2 \tr(\psibt) - \ffrac{1}{3} \nabla \cdot (\nabla \cdot \psibt) & \text{in 3D}.
 \end{cases}
}
We also proved that, in 2D, if macroscopic stresses are repulsive, then $\psi$ is a convex function. In general, these relations are insufficient to completely specify the continuum limit of this problem, just as in elasticity a missing equation is required that derives from Saint Venant's compatibility condition, via Hooke's law \cite{LandauLifshitz:1986}. 

The discrete representation of $\psi$ developed in this work allows insight into this missing stress-geometry equation \cite{TkachenkoWitten:1999,EdwardsGrinev:1999, BallBlumenfeld:2002}. Indeed, at the microscopic level, $\psi$ is only required to satisfy the virtual contact constraints, and the repulsive constraints. The former are present in all packing geometries but a very special one: the triangular lattice in 2D. On this lattice, the grains are locally and globally as densely packed as possible. In a sense, the geometry is trivial. If we assume that the repulsive constraint requires only that macroscopic stresses are repulsive, then $\psi$ can be any convex function.


In general, however, virtual contact constraints exist and couple the $\psi$ field throughout the packing. Their distribution is intimately related to the size and shape of loops. For hyperstatic packings, Newton's laws are insufficient to fully specify the stress state at the microscopic level, so we expect a family of solutions at the macroscopic level as well. These could, in principle, depend on the microscopic force law and the packing history. In the simplest case, they would depend only on the distribution of virtual contacts. 

Only at isostaticity is the microscopic stress state fully specified by the geometry. In the special case of isotropic forcing, isostaticity is achieved in a noncohesive packing when the pressure $P=0$. Using the microscopic expression \eqref{sigma}, we see that this implies $\sum_{c \in RC} |\lb^c| f_N^c =0$, where $f_N^c$ is the normal component of the contact force at $c$. For non-cohesive grains, each $f_N^c \geq 0$ and hence all normal forces vanish. Assuming a Coulomb repulsive constraint of the form
\[
\frac{1}{\mu} |f_T^c| \leq f_N^c,
\]
for any $\mu < \infty$ this implies that each tangential force vanishes as well, and hence the stress state is trivial. To understand the isostatic state for $\mu < \infty$, forces need to be renormalized. 

However, for the ideal case $\mu = \infty$, at $P=0$ only the normal components of all contact forces vanish. A nontrivial stress state is possible, consisting only of \textit{tangential} forces. From \eqref{pressure}, we see that $\nabla^2 \psi=0$ in 2D and $\nabla^2 \tr(\psibt) = \nabla \cdot (\nabla \cdot \psibt)$ in 3D. In 2D, given boundary conditions on $\psi$, this fully specifies the continuum limit. The stress state described by $\sigmab = -\nabla \nabla \psi$, with $\nabla^2 \psi=0$ is equivalent to that of a linear elastic body undergoing volume-conserving deformation.

To understand what happens at isostaticity when $\mu < \infty$, and when $P >0$, requires more sophisticated analysis. Since a noncohesive packing at isostaticity has no intrinsic force scale, by dimensional analysis, the missing equation must take the general form, in 2D, 
\[
T = P \; \mathcal{F}(\{ \rb^g \}),
\]
where $T=\sqrt{P^2-\det(\sigmab)}$ is the maximal shear stress, and $\mathcal{F}$ is an unknown function of the geometry. In a forthcoming paper \cite{DeGiulietal:2012}, we use the tools of this work to derive the missing function $\mathcal{F}$, using Edwards' statistical mechanics.

\section{Appendix A. \label{app:torque}Torque Balance in 2D} 
\begin{figure}[ht!]
  \centering
  \fbox{\includegraphics[viewport=80 70 400 300,width=0.46\textwidth,clip]{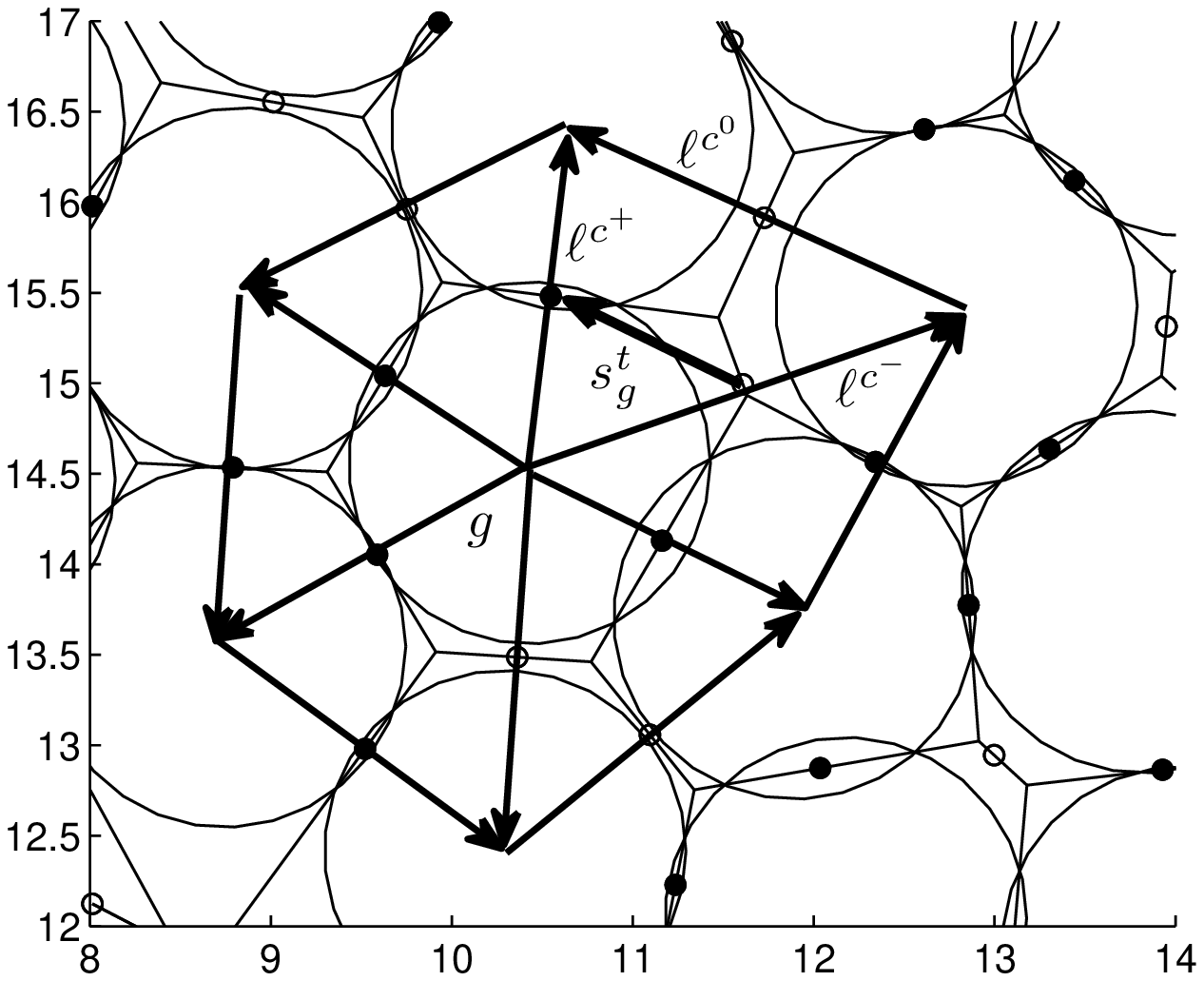}}
  \caption[Local geometry around a grain]{\label{fig:wheel} Geometry used in torque balance computation.}
\end{figure}
To verify that torque balance is satisfied identically when $\rhob = \nabla \times \psi$, we need to show that $(\nabla \cdot \rhob)^g=0$. Labelling the contacts and triangles around a grain as in Figure \ref{fig:wheel}, we note that $\bm{s}_g^{t}$ is parallel to $\lb^{ c^0(g,t)}$, so that 
\begin{align*}
A^g (\nabla \cdot \bm{\rho})^g & = \sum_{t\in T^g} \bm{s}_g^{t} \times \rhob^{t} \\
  & = \sum_{t\in T^g} \bm{s}_g^{t} \times \frac{-1}{A^t} \left( -\lb^{c^-} \psi^{c^-} + \lb^{c^+} \psi^{c^+} + \lb^{c^0} \psi^{c^0} \right) \\
  & = \sum_{t\in T^g} \frac{1}{2} (\lb^{c^+} - \lb^{c^-}) \times \frac{-1}{A^t} \left( -\lb^{c^-} \psi^{c^-} + \lb^{c^+} \psi^{c^+} \right) \\
  & = \sum_{t\in T^g} \frac{1}{A^t} \left( A^t \psi^{c^-} -A^t \psi^{c^+} \right) \\
  & = 0
\end{align*}
on summation around the grain.
\section{Appendix B. \label{app:dof}Degrees of Freedom in 3D}

Applying Euler's formula to the Voronoi polyhedron of grain $g$, we have $N_V^g - N_T^g + N_C^g = 2$. Each vertex is the confluence of three edges of the polyhedron, so that $3 N_V^g = 2 N_T^g$. Summing over the entire packing, $4N_V - N_{BV} - 3N_T + N_{BT} + 2N_C- N_{BC} = 2N$ and $12 N_V - 3 N_{BV} = 6 N_T - 2 N_{BT}$. Since the entire packing is itself a convex polyhedron with trivalent vertices, we have $N_{BV} - N_{BT} + N_{BC} = 2$ and $3 N_{BV} = 2 N_{BT}$, so that $4N_V - 3N_T + 2N_C = 2N + 2$ and $2N_V = N_T$. With these relations we can write $M_3 = 3N_T - 3N_{RG} -3(N_{VC}-N_{VG}) -3(N_V - 1)$. This shows that the loop forces $\rhob$ are constrained by torque balance and the virtual contact constraints, $N_{VG}$ of which are redundant. The gauge freedom has a dimension of $N_V-1$ because adding a constant to $\bm{B}^v$ does not change $\Delta \rhob$.

For $\psi$, we write $M_3 = N_C - 3N_{VC} + 6N_{VG} -3(N_T - N_C) - (N_C-N)- (6N_C -6 -12N_V + N)$. All but the first three terms correspond to gauge freedoms. First, we have gauge transformations of the form $\Delta \psi^c = \bm{r}^c \cdot \bm{B}$, which lead to $(\Delta \rhob)_{gc}^t = - \sgct \times \bm{B}$. Here $\bm{B}$ is a constant, but it may be derived from a fluctuating field on the triangles, viz. $\bm{B} = \sum_{t \in T^c} \bm{B}^t$. This gives $3N_T$ unknowns constrained by $3N_C$ equations. Second, we can have $\Delta \psi^c = |\bm{\ell}|^{-1} \ltildeb^c \cdot (\Delta \bm{\psi}^{g^+} - \Delta \bm{\psi}^{g^-})$, with $\Delta \bm{\psi}^{g} = \bm{r}^g B$, where again $B$ is a constant. In this case it can be derived from $B = \sum_{c \in C^g} B^c$, which gives $N_C$ unknowns constrained by $N$ equations. Gauge transformations of this type leave invariant $\rhob$. There remains a gauge subspace of dimension $6N_C -6 -12N_V + N$, the significance of which is unclear.

\section{Appendix C. \label{app:convex}Convexity of $\psi$}
The force acting on a plane with unit normal $\bm{n}$ and area $A$ is $A \bm{n} \cdot \sigmab$, with normal component $A \bm{n} \cdot \sigmab \cdot \bm{n}$. This is always positive, and hence repulsive, if and only if $\sigmab$ is a positive-definite tensor. 

In 2D, $\sigmab = \nabla \times \nabla \times \psi$ can be inverted to write $\nabla \nabla \psi = -\epsb \cdot \sigmab \cdot \epsb$. Using a matrix identity, this can be written
\begin{equation}
\label{hessianpsi}
\nabla \nabla \psi = \tr(\sigmab) \delb - \sigmab,
\end{equation}
where $\delb$ is the identity tensor. 

Writing $\bm{\hat{H}} \equiv \nabla \nabla \psi$, convexity of $\psi$ is equivalent to positive definiteness of $\bm{\hat{H}}$. The latter is equivalent to the statement that $\bm{\hat{H}}$ has positive eigenvalues. Writing $\lambda_i$ and $\bm{u}_i$ for the eigenvalues and eigenvectors of $\sigmab$, we have $\tr(\sigmab)=\sum_i \lambda_i$. From \eqref{hessianpsi}, $\bm{\hat{H}} \cdot \bm{u}_j = \sum_{i \neq j} \lambda_i \bm{u}_j$, so $\bm{u}_j$ is an eigenvector of $\bm{\hat{H}}$ with eigenvalue $\sum_{i \neq j} \lambda_i$. This is positive if $\sigmab$ is positive definite.

\section{Acknowledgements}
\begin{acknowledgments}
E.D. gratefully acknowledges discussions with Neil Balmforth, Rafi Blumenfeld, Ian Hewitt, and Christian Schoof, and NSERC for funding.
\end{acknowledgments}

\bibliography{Granular}

\end{document}